# Bribery Games on Interdependent Complex Networks

Prateek Verma[$], Anjan K. Nandi[$], and Supratim Sengupta[*]

Department of Physical Sciences, Indian Institute of Science Education and Research Kolkata,

Mohanpur – 741246, India.

$ - Equal Contribution

* - Corresponding author

E-mail: supratim.sen@iiserkol.ac.in

## Abstract

Bribe demands present a social conflict scenario where decisions have wide-ranging economic and ethical consequences. Nevertheless, such incidents occur daily in many countries across the globe. Harassment bribery constitute a significant sub-set of such bribery incidents where a government official demands a bribe for providing a service to a citizen legally entitled to it. We employ an evolutionary game-theoretic framework to analyse the evolution of corrupt and honest strategies in structured populations characterized by an interdependent complex network. The effects of changing network topology, average number of links and asymmetry in size of the citizen and officer population on the proliferation of incidents of bribery are explored. A complex network topology is found to be beneficial for the dominance of corrupt strategies over a larger region of phase space when compared with the outcome for a regular network, for equal citizen and officer population sizes. However, the extent of the advantage depends critically on the network degree and topology. A different trend is observed when there is a difference between the citizen and officer population sizes. Under those circumstances, increasing randomness of the underlying citizen network can be beneficial to the fixation of honest officers up to a certain value of the network degree. Our analysis reveals how the interplay between network topology, connectivity and strategy update rules can affect population level outcomes in such asymmetric games.

# 1. Introduction

Corruption involving bribery is an example of a social conflict scenario which affects the lives of many people particularly in developing countries. The conflict arises when a service provider (corrupt officer) withholds a service from citizens until they pay a bribe. The bribes are called harassment bribes when citizens are legally entitled to the service. Examples of such bribes include bribes paid for obtaining identification cards, electricity & gas connections, driver's license, etc. One way of discouraging such forms of corruption is by imposing a high penalty on such illegal transactions that are enforced by law. Punishment can be inflicted symmetrically on both bribe giver (who pays a bribe silently) and bribe taker (corrupt officer) or asymmetrically on only bribe taker (corrupt officer). Basu (Basu, 2011) suggested punishing only bribe takers in the case of harassment bribe will encourage citizens to act as whistle-blowers and hence would cut down incidents of bribery incidents. Several experimental and theoretical studies have been carried out to test the validity of the proposal since it was first suggested (Abbink et al., 2014; Basu et al., 2016; Dufwenberg and Spagnolo, 2015; Oak, 2015; Ryvkin and Serra, 2013; Verma and Sengupta, 2015).

The act of demanding harassment bribes showcases a conflict between a harasser (corrupt officer) and a harassed citizen. A corrupt officer would like to maximize his gain from exploiting citizens while simultaneously attempting to minimize the risk of being punished. A citizen subject to a bribe demand is faced with a different kind of economic as well as ethical dilemma. She must weigh the cost of paying a bribe against the cost of denial of service should she refuse to pay. The decisions taken in this context will depend on the bribe amount, the likelihood of redress following a complaint against the extorting officer, the cost incurred in lodging a complaint as well as other factors. The strategies adopted by the protagonists and the manner in which they are updated over time under the influence of connected neighbours has significant consequences on the prevalence and spread of such forms of corruption in society.

Analysis of such types of social conflicts using evolutionary game theory can be insightful in several ways. Firstly, such a game provides a realistic model to explore the effects of punishment on reducing corruption and enabling ethical behaviour. Several recent studies (D'Orsogna et al., 2013; D'Orsogna and Perc, 2015; Helbing et al., 2010; Lee et al., 2017, 2015; McBride et al., 2016; Rand and Nowak, 2011; Short et al., 2010; Sigmund, 2007; Szolnoki and Perc, 2013a, 2015) have begun to address the impact of punishment and reward on deterring criminal behaviour in mixed as well as structured populations. Such social conflicts characterized by competing interests and distinct strategies of interacting players also provides a natural framework for analysing evolutionary game dynamics on interdependent networks.

Initially, most studies on complex networks were focused on single networks (Barabasi and Albert, 1999; Barrat et al., 2008; Callaway et al., 2000; Cohen et al., 2000; Newman et al., 2006; Song et al., 2005; Watts and Strogatz, 1998). Networks in real life are often made up of interconnected and interdependent layers where each layer contains a network of nodes that can be different from those found in another layer (Danziger et al., 2014; Kenett et al., 2015). The interconnectedness of the two layers lead to their inter-dependence. The properties of such interacting networks can be quite different from that of a single network. A recent work on such interdependent infrastructure network reveals that small failures in a small part of one network may lead to catastrophic cascade of failures in another network (Buldyrev et al., 2010).

Evolutionary games, especially those dealing with the spread of cooperative behaviour, have been extensively analysed on a variety of networks with different topologies (Assenza et al., 2008; Fu et al., 2007; Gómez-Gardeñes et al., 2011, 2007; Kim et al., 2002; Lee et al., 2011; Masuda and Aihara, 2003; Peña et al., 2009; Poncela et al., 2011, 2007; Santos et al., 2006; Santos and Pacheco, 2005). These works established that heterogeneity in network connection positively impacts the spread of cooperation in social dilemmas though its effect is limited by payoff normalization (Masuda, 2007; Santos and Pacheco, 2006; Szolnoki et al., 2008; Tomassini et al., 2007). Recently, a number of studies have begun to focus on the conditions for spread of cooperation on interdependent networks (Gómez-Gardeñes et al., 2012; Jiang and Perc, 2013; Szolnoki and Perc, 2013b; B. Wang et al., 2012; Wang et al., 2013a, 2013b; Z. Wang et al., 2012). For example, it has been found that an intermediate interdependence between two interacting networks promotes cooperation in the population (Jiang and Perc, 2013) and unbiased coupling of payoffs of two network leads to the emergence of interdependent network reciprocity (Wang et al., 2013a). Hence properties pertaining to interdependence among players of two networks can be exploited to promote cooperation in different social dilemmas.

A common theme of all these papers is the symmetric nature of interaction among players within and across the networks with the players being identified only by their strategies. In nature, however, intra and inter network interactions are mostly asymmetric. This asymmetry may arise due to intrinsic differences in the nature of two interacting networks, environmental and genetic factors, hierarchical social structure or because players have different social roles. Asymmetric interactions can be modelled through asymmetric games also referred to in the literature as bi-matrix games (Gaunersdorfer et al., 1991; McAvoy and Hauert, 2015). The bribery game is a typical example of an asymmetric game where citizens and officers interact with each other but can only imitate strategies of members belonging to their own populations. Thus, we have one interaction graph connecting officers & citizens and two replacement graphs, one each for citizen and officer populations, specifying the connections within those populations. This framework of interdependent network interaction and dynamics can be extended to other asymmetric social conflicts as well.

In our previous work we had employed the evolutionary game theoretical model to test the validity of the hypothesis for reducing bribery incidents, proposed by Basu in a mixed (Verma and Sengupta, 2015) as well as a structured (Verma et al., 2017) population represented by a regular interdependent network. In a mixed population scenario (Verma and Sengupta, 2015) we found that imposing the asymmetric punishment scheme may not suffice in reducing incidents of harassment bribery. Factors such as how players update their strategies, the cost of complaining incurred by harassed citizens and bribe amount demanded by corrupt officers plays a significant role in determining the prevalence of corrupt officers in the population at equilibrium. Subsequently, we also found (Verma et al., 2017) that incidents of bribery can be considerably reduced in a structured population represented by a regular inter-dependent network, in comparison to the mixed population scenario. Another key feature of such networks was the optimal range of connectivity of the nodes in the citizen and interaction networks that facilitated fixation of honest officers. While regular networks offer some insight into the role of underlying population structure and connectivity, real-world social networks are more complex and described by either small-world or scale-free networks. The main motivation of the current work is to understand how heterogeneity in network connections of both citizen and officer networks (implying differences in influencing individual behaviour by connected neighbours) together with asymmetry in the citizen and officer population sizes impact the spread of honest officers in the asymmetric penalty scenario. As in the previous work, we have focused primarily on varying two important parameters of the bribery game namely, bribe amount ($b$) demanded by the corrupt officer and cost of complaining ($t$) incurred by complaining citizens. This was based on the grounds that these two parameters are more easily controlled than the others. Parameters such as amount of punishment imposed and prosecution rates can vary across states and even jurisdictions and changes are more likely to depend on fickle political processes. Nevertheless, we have also analysed the effect of varying bribe amount ($b$) and punishment ($p_o$) on the equilibrium distribution of strategies in the population.

We find that a change in network topology from regular to small-world to random adversely affects the spread of honest officers in certain regions of parameter space. However, the extent of the decrease in dominance of honest officers and complaining citizens depends on the rewiring probability and the average number of links per citizen in the network. As in the case of regular networks, reduction in incidents of bribery over the parameter space is maximized for an optimal value of citizen network degree. However, this no longer holds for a completely random network.

The behaviour is reversed if an asymmetry exists in the population sizes of the citizens and officers. In this case, the success of honest officers is maximized when the network degree is minimum with an initial increase in degree leading to a sharp drop in the fixation of honest officers. Subsequent increase in the degree of the citizen network aids in the fixation of honest officers. Moreover, increasing randomness of the citizen network encourages the spread of honest officers.

## 2. Methods

### 2.1 Bribery game and update rule

We start with two sets of population, one of officers and the other of citizens as shown in the schematic Fig. 1. Officers are connected with citizens through a regular interaction network (network 3) of degree $IN$ and with other officers through an intra-officer network (network 1) of average degree $ON$. Citizens, similarly are connected among themselves through intra-citizen network (network 2) of average degree $CN$. $IN$ defines the number of officer(s) each citizen interacts with. The interaction network specifies how citizens play the bribery game with officers while the intra-citizen and intra-officer networks determine how a citizen and an officer updates their respective strategies by consulting connected neighbours in their corresponding networks. Depending upon their own strategies as well as strategies of opponents, players receive payoffs. The payoff matrix for the bribery game is given by:

$$M_1 = \begin{array}{c} \\ C_1 \\ C_2 \\ C_3 \end{array} \begin{pmatrix} O_1 & O_2 \\ c & c-b \\ c & c-b-t+k(r-p_c) \\ c & -t \end{pmatrix}$$

$$M_2 = \begin{array}{c} \\ O_1 \\ O_2 \end{array} \begin{pmatrix} C_1 & C_2 & C_3 \\ v & v & v \\ v+b & v+b-k(p_o+r) & v-kp_o \end{pmatrix}$$

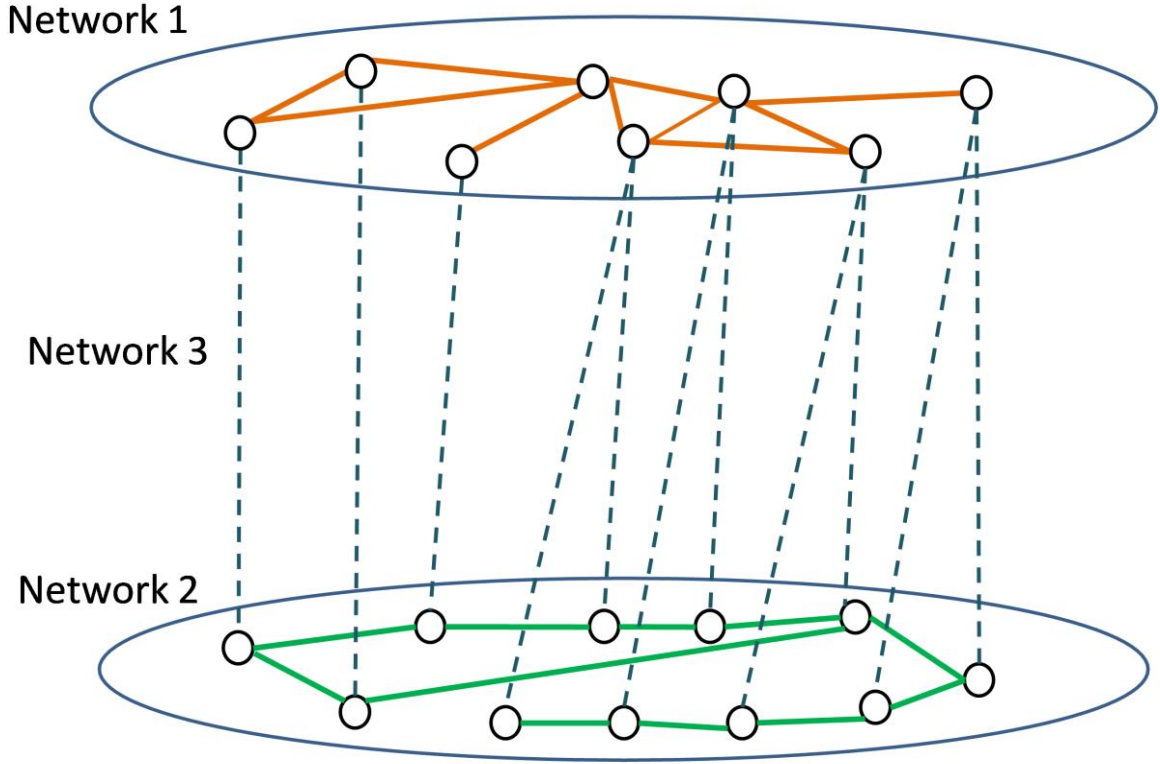

Fig. 1: Schematic diagram of interdependent network system of evolutionary game. Players in network 1 and 2 denoted by circles are connected with orange and green links respectively. Players of network 1 and 2 are connected through inter-network 3 represented by blue dashed links. Players of both population are also connected among themselves and update their strategies via intra-network (network 1 and 2). Topologies of network 1, 2 and 3 plays an important role in deciding the evolutionary fate of the population.

When a citizen irrespective of her strategy interacts with an honest officer ($O_1$), the officer and citizen receive a fixed payoff of $c$ and $v$ respectively. If the officer is corrupt ($O_2$), payoff depends on the strategy of citizen. Citizens who pay the bribe amount $b$ silently to the corrupt officer are apathetic citizens ($C_1$). Those who pay the bribe and complain to the concerned authorities are conscientious citizens ($C_2$). Citizens who refuse to pay the bribe and then complain are called honest citizens ($C_3$). Complaining citizens ($C_2$ and $C_3$) also have to bear the cost of complaining $t$. Not all complaints lead to the prosecution of corrupt officers, it happens with the probability of prosecution $k$. When prosecuted, $O_2$ and $C_2$ must pay a penalty amounting to $p_o$ and $p_c$ respectively. $O_2$ is punished for taking the bribe and $C_2$ is punished for giving a bribe. In the asymmetric penalty scenario, when citizens are not liable for giving bribes, no punishment is levied on the citizens ($p_c = 0$). $r$ is the refund $C_2$ gets when $O_2$ is prosecuted for taking a bribe, which happens with probability $k$. All our results are obtained for the asymmetric liability with refund ($r = b$) scenario.

If $IN > 1$, an officer plays the game with $IN$ citizens per round. After bribery game is played between citizens and officers via the interaction network, the payoffs of each of the officers are averaged (over $IN$ interactions with citizens) and strategies of both citizens and officers are updated. The strategy update happens within the networks of respective population via a synchronous update process in which the entire population is updated simultaneously following the interactions. The player chosen for updating her strategy is called the 'focal player'. Focal player compares her payoff with the payoff of one of a randomly chosen neighbour known as 'role model'. If the payoff of focal player ($\pi_f$) is greater than or equal to that of the role model ($\pi_r$) then focal player retains her original strategy. If $\pi_r > \pi_f$, the focal individual imitates the role model's strategy with a probability given by:

$$\frac{\pi_r - \pi_f}{\max(\Delta \pi)}$$

where $\max(\Delta \pi)$ is the maximum possible payoff difference between any two players belonging to the same population. The process of interaction and population update is continued until the code has run for a fixed number of generations or until one of the officers' strategies gets fixed in the population, whichever is earlier.

'Small-world' networks represent a more realistic model of social networks and are marked by their short characteristic path length as compared to the regular networks. This implies that any individual node can be reached from any other node by traversing a small number of links. Such a network can be formed by rewiring a regular network via Watts-Strogatz algorithm where each link from the node is removed from one end with probability $p$ and reconnected to a node chosen uniformly at random (avoiding self-loops and link duplication). Rewiring each link of a regular network allows the rewiring probability $p$ to be used as a tuneable parameter that allows the network topology to change from a regular ($p = 0$) to completely random 'Erdos-Renyi' network ($p = 1$). We did not allow the network to evolve in a manner that results in more than one connected component.

Small-world character of a network is obtained for a low value of $p$, typically around the value of 0.10 (Watts and Strogatz, 1998) subject to the additional criteria $N_C \gg CN > \ln(N_C)$. This condition is satisfied in our case for $CN < 10$ when $N_C = 100$ and $CN < 100$ when $N_C = 1000$. Nevertheless, we varied $CN$ across a wide range since we wanted to study the effect of varying network connectivity as the network topology changes from a regular to a completely random one. We ensured that the evolution of the network because of rewiring did not result in a network with more than one connected component. We also examined the effect of scale-free network topology (Barabasi and Albert, 1999) on the prevalence of bribery by constructing scale-free citizen network following the Barabasi-Albert method of network growth via preferential attachment. We start with a fixed number ($m_0$) of all connected nodes $m_0$. New nodes are linked to $m$ ($<m_0$) existing nodes with probability proportional to the degree of existing nodes. In this way, new nodes are more likely to attach to nodes of higher degree than nodes with a lower degree. The number of new links ($m$) that are connected to existing nodes decide the average degree per node of the network. The network topology did not change during the course of the game even though the nodes themselves, representing strategies employed by a citizen and officer could change during the strategy update process.

When rewiring probability $p$ increases, a close look at the equilibrium distribution reveals that for a few values of $b$ & $t$, honest and corrupt officers coexist at the equilibrium although the fraction of one of these two strategies is much less than the other. Such coexistences can be attributed to be artefacts of the rewiring process. It is possible that the rewiring of a connection between the pair of nearest-neighbour citizens that define a cluster boundary leaves the two clusters unconnected. Such a structure can lead to the coexistence of $O_1$ and $O_2$ officers because a single such officer is unable to change her strategy. When the probability of rewiring increases, chances of having network structures favouring such coexistence increases. Therefore, the region of parameter space where $O_1$ and $O_2$ officers coexisted increases with rewiring probability. When we allow each officer to flip her strategy with a small mutation probability $\mu = 10^{-4}$ at each time step, such artefacts disappear, and the equilibrium concentrations exhibit no coexistence of $O_1$ and $O_2$ officers. The strategy flipping probability ($\mu$) is much smaller than the strategy update probability and hence does not affect the update dynamics in any way. We used this value of $e$ to generate all figures.

## 3. Results and Discussion

In this and subsequent sections, we present results to show the effect of rewiring of the citizen network, officer network, variations in citizen network connectivity, asymmetry in the size of citizen and officer populations on the prevalence of bribery. Since scale-free networks provide an alternative mode of constructing social networks, we also examine how the scale-free network topology of the citizen population affects the spread of corrupt officers.

## 3.1 Effect of rewiring the citizen's network

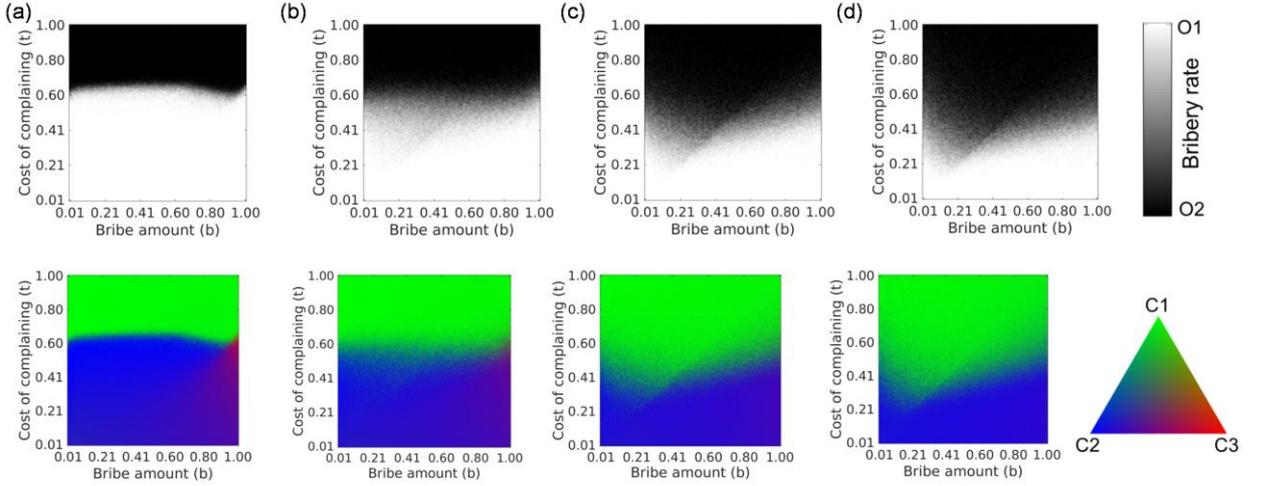

Fig. 2: Equilibrium distribution (averaged over 100 realizations) of officers (upper panel) and citizens (lower panel) with increasing rewiring probability of citizen's network and small mutation ( $\mu = 10^{-4}$ ) in officer's population for (a) $p = 0$ (regular) (b) $p = 0.1$ (c) $p = 0.4$ and (d) $p = 1.0$ (completely random). The colours in the upper panel represent the fraction of corrupt officers ($O_2$) while those in the lower panel represent the fraction of citizens employing one of the three possible strategies. Black indicates fraction of $O_2$ is 1. Green, blue and red colors represent points where the fraction of citizens who pay silently ( $C_1$ ), pay & complain ( $C_2$ ) and refuse to pay ( $C_3$ ) respectively, is 1. Fixed parameter in the simulation: $N_C = 100$, $N_O = 100$, $IN = 1$, $CN = 2$, $ON = 2$, $v = 1$, $c = 1$, $k = 0.4$, $p_o = 2$, $p_c = 0$, $r = b$.

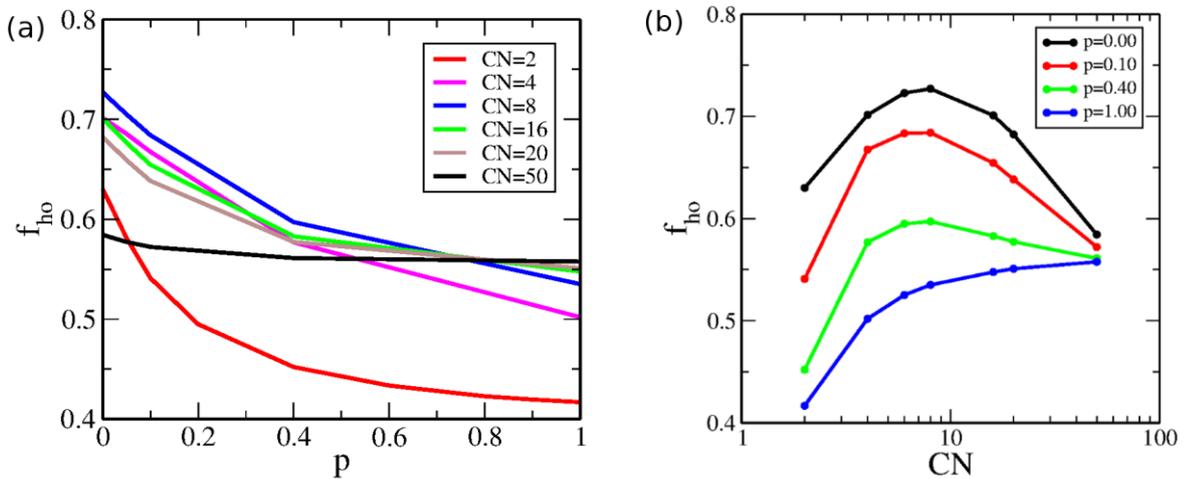

Fig. 3: Fraction of phase space $f_{ho}$ (averaged over 100 realizations) for which honest officers can get

fixed is plotted against (a) rewiring probability of citizen's network $p$ for different values of $CN$. Increasing randomness of citizen's network negatively impacts the spread of honest officers over phase space. (b) $f_{ho}$ versus $CN$ for different values of rewiring probability $p$. Other parameters are fixed to $N_C = 100$, $N_O = 100$, $IN = 1$, $CN = 2$, $ON = 2$, $\mu = 10^{-4}$, $v = 1$, $c = 1$, $k = 0.4$, $p_o = 2$, $p_c = 0$, $r = b$.

Initially we chose to rewire the citizen's network only since it is more likely to have a complex network topology due to each citizen forming a variety of connections with friends and relatives, thereby being in a position to influence and be influenced by their connected neighbours. The service providing officers on the other hand are more likely to be influenced by connected counterparts in the same office and their chances of forming long-range connections with other officers providing the same service are relatively low. Hence it seems reasonable to consider the officer network to be a regular one in contrast to the more complex topology of the citizen network. We start with symmetric populations of size $N_O = N_C = 100$ with $ON = CN = 2$ and $IN = 1$. Fig.2 shows the equilibrium concentration of different strategies for varying bribe amount $b$ and cost of complaining $t$. In Fig. 2(a) ($p = 0$) for a high cost of complaining ($t \sim 2/3$ or more), both corrupt officers and corresponding apathetic citizens get fixed in the population. On the other hand, for low values of $t$, the spread of honest officers in the population is accompanied by increase in the concentration of conscientious and honest citizens. For high bribe amount and low cost of complaining, honest citizens play an important role in turning the tide towards the fixation on honest officers. As the citizen network topology changes from regular to small-world, honest officers are found to prevail only for a lower cost of complaining except when bribe demanded is very high. In that regime, the presence of honest citizens who refuse to pay the bribe ($C_3$) tips the balance in favour of honest officers (lower panels of Fig. 2a, b). For $CN = 2$, with further increase in rewiring probability of the citizen network, the region of $b - t$ phase space over which honest officers can get fixed ($f_{ho}$) decreases (Fig.3a). Hence increasing the randomness of citizen network negatively impacts the fixation of honest officers in the population. Fig.4(a) and (b) shows the effect of going from a small-world ($p = 0.1$) to a completely random ($p = 1$) network on the population dynamics for a specific parameter set. Even though strategies were randomly distributed across the initial population of citizens and officers, we find interacting clusters forming very quickly between apathetic citizens ($C_1$) & corrupt officers ($O_2$) and complaining citizens ($C_2$ and $C_3$) & honest officers ($O_1$). However, the strategy update dynamics leads to distinct outcomes in the two cases with corrupt officers prevailing when the citizen network is random. Movies showing the update dynamics that eventually leads to fixation of honest and corrupt officers for $p=0.1$, $CN=2$ (Movie 1a) and $p=1$, $CN=2$ (Movie 1b), $p=0.1$, $CN=20$ (Movie 1c) and $p=1$, $CN=20$ (Movie 1d) respectively for small ($CN=2$) and large ($CN=20$) values of the citizen network degree are available as Supplementary Material.

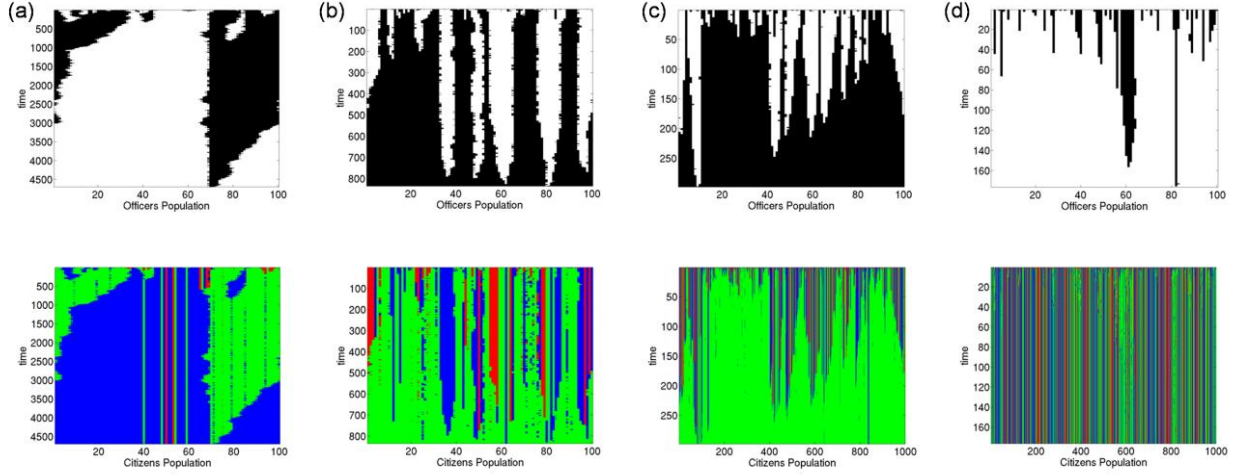

Fig. 4: Time evolution of strategies employed by officers (upper panel) and citizens (lower panel) at fixed values of bribe amount ($b$), cost of complaining ($t$) for symmetric (a & b) and asymmetric (c & d) citizen populations sizes for two different rewiring probabilities. (a) $N_C=100$, $N_O=100$, $p=0.1$ (b) $N_C=100$, $N_O=100$, $p=1.0$ (c) $N_C=1000$, $N_O=100$, $p=0.1$ (d) $N_C=1000$, $N_O=100$, $p=1.0$. Black and white colors represent corrupt and honest officers respectively. Green, blue and red colors represent citizens who pay silently ($C_1$), pay & complain ($C_2$) and refuse to pay ($C_3$) respectively. Other parameters are: $IN=1$, $ON=2$, $v=1$, $c=1$, $b=0.2$, $t=0.5$, $k=0.4$, $p_o=2$, $p_c=0$, $r=b$.

An analysis of the strategy-update dynamics suggests a plausible qualitative explanation for these results. Usually after a few successive updates, corrupt officers ($O_2$) form interacting clusters with apathetic citizens ($C_1$) and honest officers ($O_1$) form interacting clusters with complaining citizens ($C_2$ and $C_3$). The strategy-update dynamics at the cluster boundary determines the growth of one and corresponding shrinkage of the other cluster, and such processes can lead to the eventual fixation of a specific strategy in the population over time. In regular networks, the strategy of individuals at the boundary of a cluster can influence those of their connected neighbours in the adjacent cluster. This makes the spread of honest officers and complaining citizens favourable under certain conditions. However, an increase in rewiring probability which leads to the establishment of long-range connections (at the cost of near-neighbour connections) in the citizen network can reduce the ability of individuals at a cluster boundary to influence counterparts in the adjacent cluster because they may no longer be connected neighbours after the rewiring process. This effect is prominent for low to moderate values of $CN$. As $CN$ increases, the decrease in $f_{ho}$ becomes less pronounced (see coloured lines in Fig. 3a) and eventually disappears for very large $CN$ (black line Fig. 3a) when the degree of the citizens network is large enough to influence neighbours lying in different clusters. The detrimental effects of increasing randomness of the citizen network is offset by a large network degree that ensures majority of the nodes are connected.

The average degree of the citizen network has a significant effect on fraction of the $b-t$ parameter space for which honest officers get fixed in the population. Supplementary Figures 1 and 2 depicts the equilibrium concentrations of different strategies in both the citizen and officer networks for $p=0.1$ and $p=1$ respectively as $CN$ is varied. It is clear from Supplementary Fig. 1 that there exists an optimal value of $CN$ which facilitates fixation of honest officers. This is summarized in Fig. 3b where the variation of $f_{ho}$ with the average network degree ($CN$) is plotted for different values of $p$. We observe that the area of $b-t$ space over which honest officers dominate is highest for $CN=6$ (Fig. 3b). When the citizen network becomes completely random (see Supplementary Fig. 2), no such optimality with $CN$ is observed. An increase in $CN$ leads to a rapid initial increase in $f_{ho}$ up to a point beyond which $f_{ho}$ nearly saturates with further increase in $CN$ since the advantage of increasing $CN$ follows the law of diminishing returns. The blue line in Fig. 3b clearly depicts this trend. As the degree of the citizen network approaches the size of the network, the effect of increasing randomness of the network on $f_{ho}$ gradually diminishes as can be seen from the clustering of $f_{ho}$ values for different $p$ in Fig.3b

Since increasing the randomness of citizen's network facilitates the spread of corrupt officers (Fig. 2) when the officer network was regular, we also wanted to examine the effect of simultaneously increasing the randomness of both the citizen and officer networks. When both networks are rewired with same probability ($p'$), the area of parameter space over which honest officers get fixed decreases significantly compared to the case when only the citizen network is rewired as shown in Supplementary Fig. 3. Rewiring of the officer network is detrimental to the spread of honest officers for the same reason as the rewiring of citizen network.

### 3.2 Asymmetric population size
In realistic scenarios, an officer providing a service typically interacts with many citizens seeking the service. Hence the citizen population is expected to be larger than the officer population. To understand the effect of *asymmetric* population size we carried out simulations where the citizen's population was 10-100 times larger than the officer's population. This means there are 10-100 citizens connected to each officer through the interaction network. However, each citizen is still connected to $IN$ officers and the larger size of the citizen population does not change $IN$. The *b-t* diagram of equilibrium frequencies of officer and citizens shows that increasing the randomness of citizen's network increases the parameter space for honest officer to dominate (Fig.5), with the increase coming primarily from the region of parameter space characterized by low bribe-amount.

Fig.4(c) and (d) depicts the evolution of officer and citizen strategies over time as the degree of randomness changes from ($p=0.1$) to a completely random ($p=1$) network and shows that increasing randomness favours fixation of honest officers for the specified parameter set. This trend is the opposite of what we observed for the symmetric population case (Fig.2), though it is worth emphasizing the absolute value of $f_{ho}$ was much higher across the entire range of $p$ values in the latter case as can be observed by comparing the black line with coloured lines in Fig. 6a. Fig. 4(c) and (d) further reveal a strong correlation between the dominance of apathetic citizens within the blocks of 10 citizens that interact with a single officer and the eventual fixation of corrupt officers. In Fig. 4(d), the strategies of citizens within a block are more heterogeneously distributed. This prevents corrupt officers from increasing their payoff by exploiting a large number of apathetic citizens within a block and inhibits their spread in the officer population.

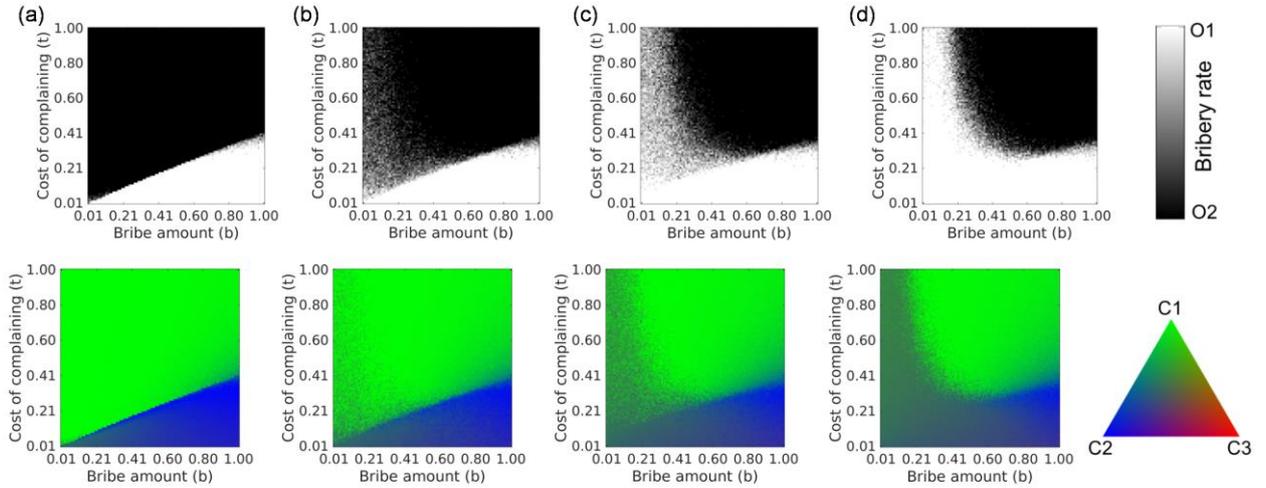

Fig.5: Equilibrium distribution (averaged over 10 realizations) of officers (upper panel) and citizens (lower panel) for increasing rewiring probability of citizen's network and asymmetric population with parameters $N_C=1000$, $N_O=100$, $IN=1$, $CN=2$, $ON=2$ (a) $p=0$ (b) $p=0.1$ (c) $p=0.4$ and (d) $p=1.0$. Other fixed parameters are $\mu=10^{-4}$, $v=1$, $c=1$, $k=0.4$, $p_o=2$, $p_c=0$, $r=b$.

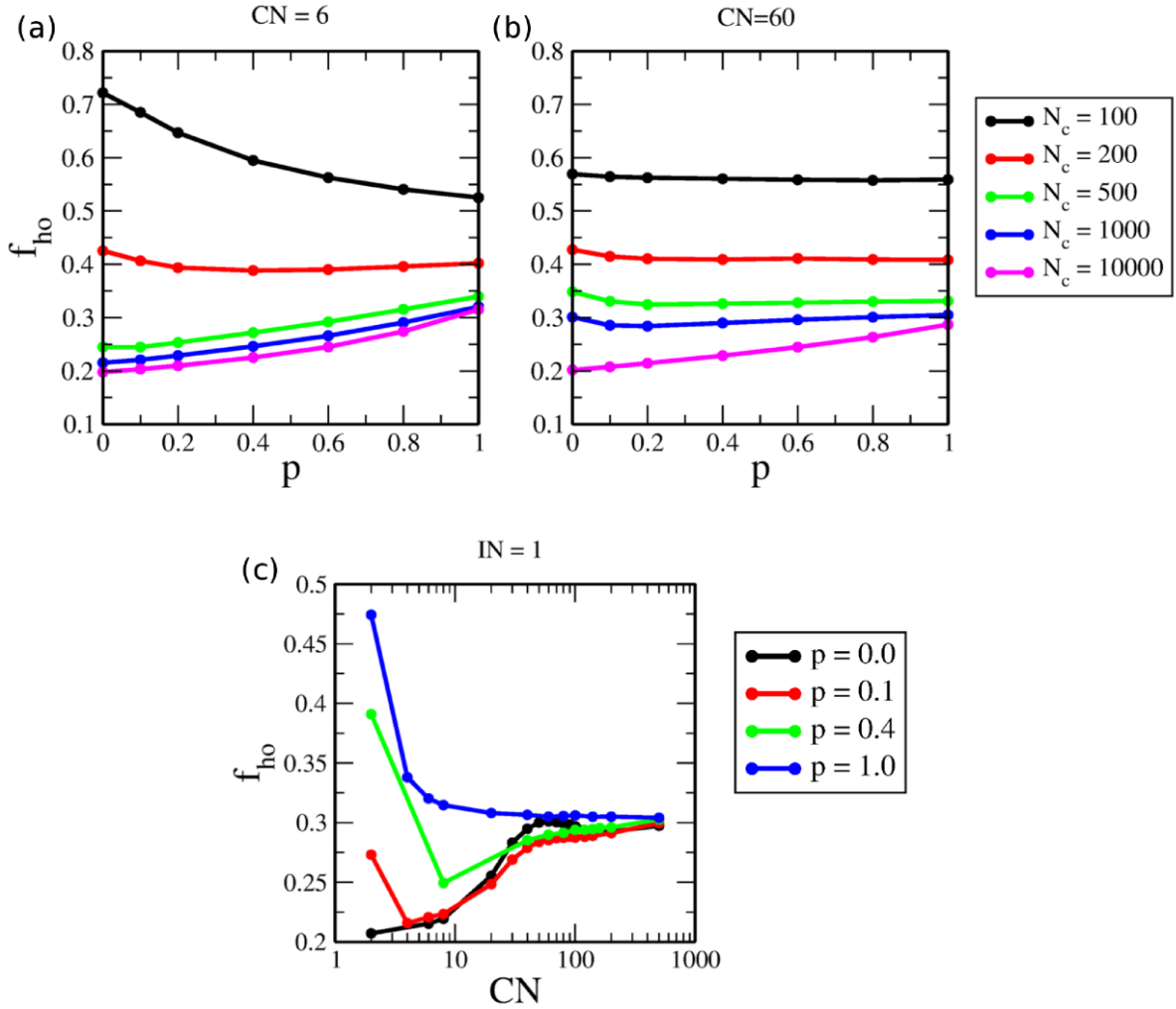

Fig. 6: Fraction of phase space $f_{ho}$ (averaged over 100 realizations) for which honest officers can get fixed is plotted against (a) rewiring probability of citizen's network $p$ for different values of $N_C$ for (a) $CN = 6$ (b) $CN = 60$ (c) $f_{ho}$ as function of $CN$ for different $p$ when $N_C = 1000$. Other parameters are fixed to $N_O = 100$, $IN = 1$, $ON = 2$, $\mu = 10^{-4}$, $v = 1$, $c = 1$, $k = 0.4$, $p_o = 2$, $p_c = 0$, $r = b$.

When $CN$ is small and a group of 10 citizens interact with the same officer, the strategies of citizens within the block of 10 are less likely to be influenced by strategies of citizens in an adjacent block when $p = 0$ (regular network). Complaining citizens in the interior of the 10-citizen block when interacting with a corrupt officer receives a lower payoff and is therefore more likely to imitate their apathetic neighbours within the block. Change in strategies of such citizens are controlled by change in the corresponding officer's strategy. However, when $p$ increases, a citizen within a block can acquire connections with citizens in other blocks. She can therefore influence and be influenced by those citizens as well, and not just by their connected neighbours within a block. This influence of their connected neighbours in other blocks reduces the ability of the corresponding officer to control the citizens' strategy through their interaction. This leads to faster variation in citizen's strategies within a block and is beneficial for the spread of honest officers. This advantage exists as long as the degree of the citizen's network is approximately less than the block size. As the degree $CN$ increases, the ability of each citizen to influence interacting partners in other blocks increases. Thus, increasing $p$ provides no additional advantage to honest officers and this is seen by comparing the variation of the green and blue lines with $p$ in Fig. 6a ($CN = 6$) and Fig. 6b ($CN = 60$). In the former case, a slight increase in $f_{ho}$ is observed on changing the citizen network from regular to random. In contrast, in the latter case, $f_{ho}$ remains unchanged on varying the rewiring probability unless $N_C$ is increased to 10,000 which corresponds to the situation where a group of 100 citizens interact with a single officer. Fig. 6c shows the variation of $f_{ho}$ with the degree of the citizen network for a fixed ratio ($N_C / N_O = 10$) of citizen to officer population sizes. A contrast with the symmetric population case is noteworthy here. In that case, we found an optimal value of $CN$ which maximizes $f_{ho}$ (see Fig. 2). In the asymmetric population case, this is observed only for $p = 0$. As $p$ increases, the peak vanishes and for non-zero rewiring probability, an initial increase in $CN$ leads to a *decrease* in $f_{ho}$ with the extent of the drop depending on the degree of randomness of the citizen network. Subsequent increase in $CN$ leads to a gradual increase in $f_{ho}$ which eventually saturates at a value that is independent of the rewiring probability of the citizen network.

In Fig. 6c ($0 < p < 1$), an increase in $CN$ increases intra-block connections, facilitating the spread of apathetic citizens ($C_1$) since payoff of $C_1$ is greater than $C_2$ and $C_3$ when playing against corrupt officers ($O_2$). If number of complaining citizens within the block are high, payoff of $O_1$ would be greater than $O_2$ hence $O_2$ would not spread. However, when the number of apathetic citizens ($C_1$) becomes sufficiently large within the block, corrupt officer gets stabilised ($\pi_{O2} > \pi_{O1}$) and can replace $O_1$. Initially, an increase in $CN$ can only increase intra-block connection to the extent where all citizens within the block are connected; therefore $f_{ho}$ reaches a minimum when $CN$ approaches number of citizens within a block ($N_C / N_O = 10$ in our case). Further increase in $CN$ increases the adjacent inter-block connections which does not give enough time for $O_2$ to get stabilised hence $O_1$ can spread more easily ($\pi_{C2}, \pi_{C3}$ against $O_1 > \pi_{C1}, \pi_{C2}, \pi_{C3}$ against $O_2$) leading to the increase in $f_{ho}$ from its minimum value. The advantage of inter-block connections in facilitating the spread of honest officer diminishes when the citizen network approaches a complete graph (where all citizens are connected with every other citizen) hence $f_{ho}$ eventually saturates for very high $CN$ (Fig. 6c).

We also analysed the consequence of increasing the rewiring probability (Supplementary Fig.4) and citizen network degree (Supplementary Fig.5) in the parameter space obtained by varying punishment ($p_o$) and bribe amount ($b$). In contrast with the mixed population scenario, increasing the penalty on corrupt officials makes no positive impact for low bribe amounts in populations characterized by small-world networks. As the randomness of the underlying social network increases, a high penalty increases the likelihood of fixation of honest officers due to the increase in number of complaining citizens. A similar trend in observed as the citizen network degree is increased for the underlying small-world network characterized by $p = 0.1$

**3.3 Scale-free network**

Many real networks possess the scale-free property characterized by a power law degree distribution. In such networks, a small number of individuals (labelled as hubs) have a large number of links while most of the others have a relatively small number of connections. By their very nature, the hubs have the ability to influence and be influenced by a much larger set of connected citizen neighbours compared to other individuals in the network. We examined the effect of varying the average citizen network degree ($CN$) on the prevalence of bribery. Fig. 7 shows the phase plot obtained for scale-free citizen's network with different average degree. Increase in $CN$ decreases the area of the parameter space over which honest officers can prevail. The change is especially significant in the low-$b$, moderate to high-$t$ region of parameter space. This trend is quite similar to the result obtained in for the random network case (see blue curve in Fig. 6c). Scale-free networks are different from random networks because of the presence of a small number of 'hubs' (nodes with a higher degree) but they are also similar in some respects. In the scale-free network, citizens within the blocks are more likely to be connected with citizens outside their blocks. Greater inter-block connections facilitate the spread of honest officers in the population like in case of the completely random network ($p = 1$).Increasing the average degree $CN$ leads to increase in the number of intra-block connections (links within the block) which helps in the spread of apathetic citizens and corrupt officers (indirectly) resulting in the decrease of $f_{ho}$. When the average degree ($CN$) approaches the population size, the network tends towards one described by a complete graph. Hence further increase in $CN$ has little effect on the value of $f_{ho}$. The effect of greater heterogeneity in the degree of the scale-free network is also diluted since the payoffs of individual citizens are averaged over number of connected neighbors before each population update via imitation.

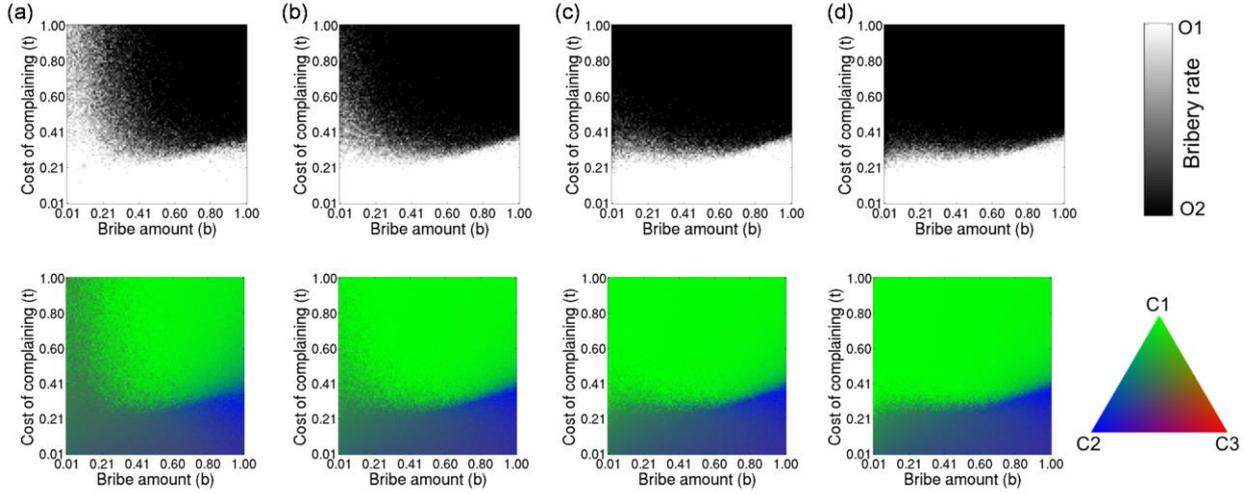

Fig.7: Equilibrium distribution (averaged over 100 realizations) for officers (upper panel) and citizens (lower panel) with increasing $CN$ for scale-free citizen's network (constructed via Barabasi-Albert algorithm) with parameter $N_C = 1000$, $N_O = 100$, $IN = 1$, $ON = 2$, initial number of nodes: $m_0 = m+1$, (a) $CN = 2$, $m = 1$ (b) $CN = 4$, $m = 2$ (c) $CN = 8$, $m = 4$ and (d) $CN = 40$, $m = 20$. Other fixed parameters in the simulation: $\mu = 10^{-4}$, $v = 1$, $c = 1$, $k = 0.4$, $p_o = 2$, $p_c = 0$, $r = b$.

## 4. Conclusions

Our work provides important insights into collective behavioural outcomes in social conflicts and reveals how those outcomes depend on the underlying organization of the population. Our results show how changes in individual decisions that are affected by the nature of the social network can lead to emergence of population-level outcomes that could not be predicted on the basis of individual interactions only. The underlying social structure and the number of connections each individual possesses, plays a key role in determining how individual behaviours change under the influence of connected neighbours and how the effect of such changes percolate through the population.

For identical citizen and officer population sizes, the region of parameter space for which honest officers can get fixed in the population decreases when citizen network's topology changes from regular to random. Making citizen's network more random increases the likelihood of severing of near-neighbour links and establishment of long-range links between citizens. This limits a citizen's ability to influence other citizens and hence other officers in her vicinity thereby hindering the growth of clusters of honest officers. Increasing the average number of links per node ($CN$) for the citizen network ensures increased near neighbour connections and offsets the negative effect of increasing network randomness leading to a slower decrease in $f_{ho}$ that eventually flattens out as the network degree approaches the network size. Random rewiring of links does not have much effect on an almost completely connected citizen network. As in the case of regular network (Verma et al., 2017), a peak in $f_{ho}$ for an optimal network degree is also observed as long as the network topology is not completely random (i.e. $p<1$).

When the number of service seekers are larger in comparison to service providers, more citizens visit the same officer to access the service resulting in the interaction of a block of citizens with a single officer. In general, an increasing asymmetry (implying larger citizen block size per officer) between the citizen and officer population sizes facilitates spread of corrupt officers over a larger region of parameter space for all complex network topologies obtained by varying the rewiring probability. However, the effect of varying rewiring probability on $f_{ho}$ depends on the network degree. When the degree exceeds the block size ($N_C/N_O$), the ability of citizens to influence their counterparts in neighbouring blocks nullifies the effect of increasing rewiring probability on $f_{ho}$. On the other hand, when the degree is less than or comparable to the block size, an honest citizen's influence is mostly restricted within a block. Links across blocks becomes more important than links within the blocks for the spread of honest officers in this scenario. Increasing the rewiring probability facilitates the establishment of long-range inter-block links and thereby aids the expansion of clusters of honest citizens with a resulting increase in $f_{ho}$. Intriguingly, for a non-zero rewiring probability, increasing $CN$ initially has a detrimental effect on the spread of honest officers in some regions of parameter space. Such an increase ensures increased connectivity of citizens within a block. The payoff structure then helps in faster spreading of apathetic citizens within a block. This has a cascading effect by first increasing the payoff of the corrupt officer which in turn leads to the rapid spread of corrupt officers. Subsequently, this effect is modulated when further increase in $CN$ establishes links across blocks and allows citizens to influence counterparts in other blocks.

The results presented here indicates how the intricate interplay between interaction across networks, rules for changing individual behaviour and network connectivity lead to complex evolutionary dynamics of the competing strategies. The outcome is manifest through shifting phase boundaries in parameter space. Our analysis while specific to bribery games can be easily generalized to any symmetric or asymmetric games. Exploring the effect of changing the topology of the interaction network and the consequences of introducing an overlap between the interaction and update networks are exciting directions for future explorations. Such investigations will lead to improved understanding of the complex dynamics of cooperation and conflict that underscore evolutionary games on interdependent networks.

Our results for complex interdependent networks indicate that in contrast to regular networks, corrupt officers and apathetic citizens proliferate more rapidly when the bribe-amount demanded is low and cost of complaining is moderate to high. Reducing the cost of complaining would be an effective means of reducing incidents of bribery but the extent of reduction necessary, depends non-trivially on the social organization including the average number of connections each citizen possesses. Hence, policies designed to reduce corruption need to take into account the underlying social structure in order to be effective. We believe our work can inform more effective policy-making to combat harassment bribery.

## Acknowledgement

P.V. is funded through an INSPIRE graduate fellowship provided by DST, India.

## Supplementary Figures

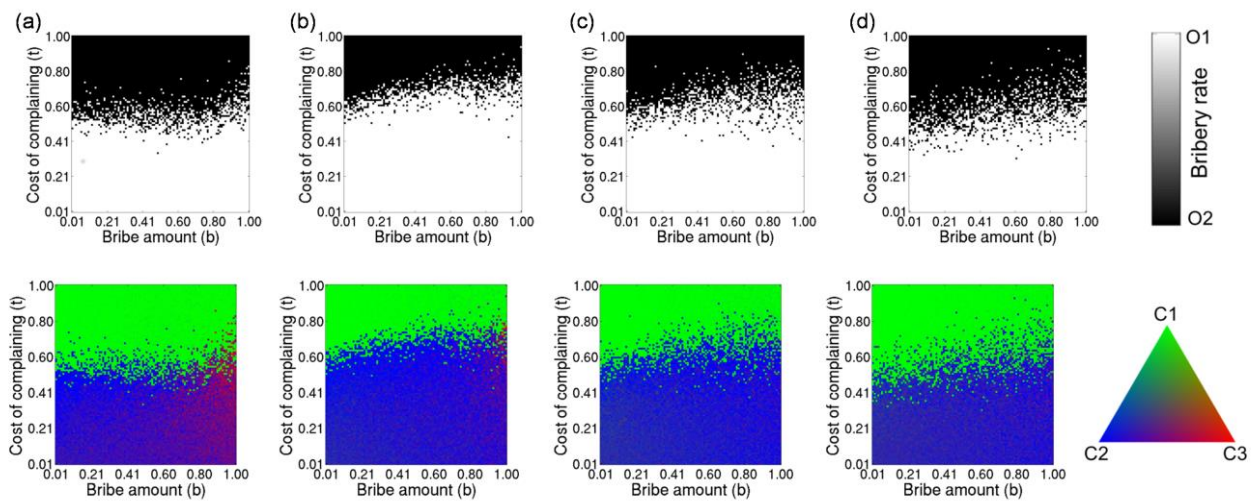

Supplementary Fig. 1: Equilibrium distribution for officers (upper panel) and citizens (lower panel) with increasing $CN$ with parameters $N_C = 100$, $N_O = 100$, $IN = 1$, $ON = 2$ and $p = 0.1$ for (a) $CN = 2$ (b) $CN = 6$ (c) $CN = 20$ and (d) $CN = 50$. Other fixed parameters in the simulation: $v = 1$, $c = 1$, $k = 0.4$, $p_o = 2$, $p_c = 0$, $r = b$, $\mu = 10^{-4}$.

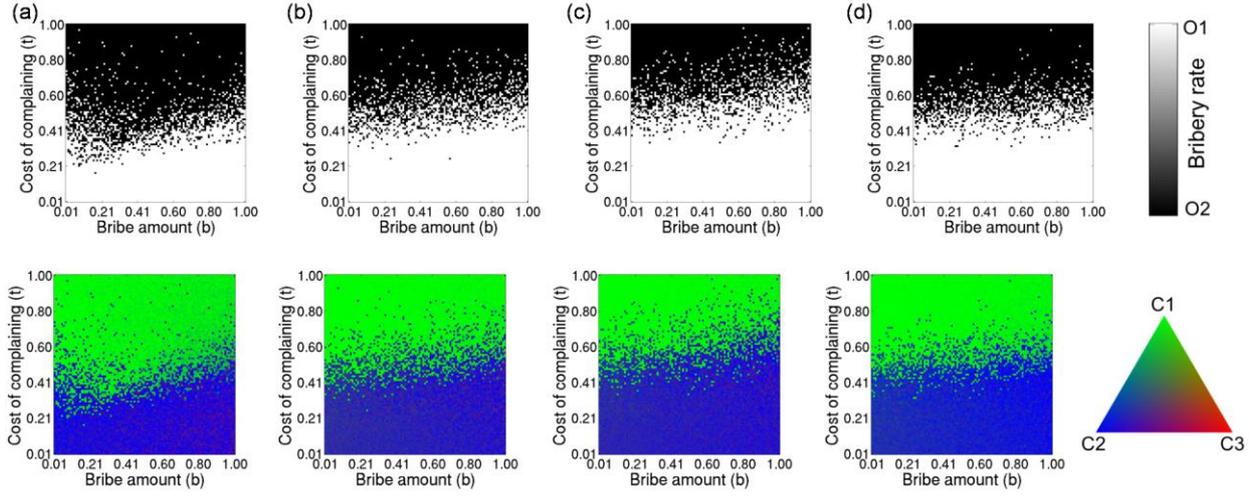

Supplementary Fig. 2: Equilibrium distribution for officers (upper panel) and citizens (lower panel) with increasing $CN$ with parameter $N_C = 100$, $N_O = 100$, $IN = 1$, $ON = 2$, and $p = 1$ (citizen network is completely random) for (a) $CN = 2$ (b) $CN = 6$ (c) $CN = 20$ and (d) $CN = 50$. Other fixed parameters in the simulation: $v = 1$, $c = 1$, $k = 0.4$, $p_o = 2$, $p_c = 0$, $r = b$, $\mu = 10^{-4}$.

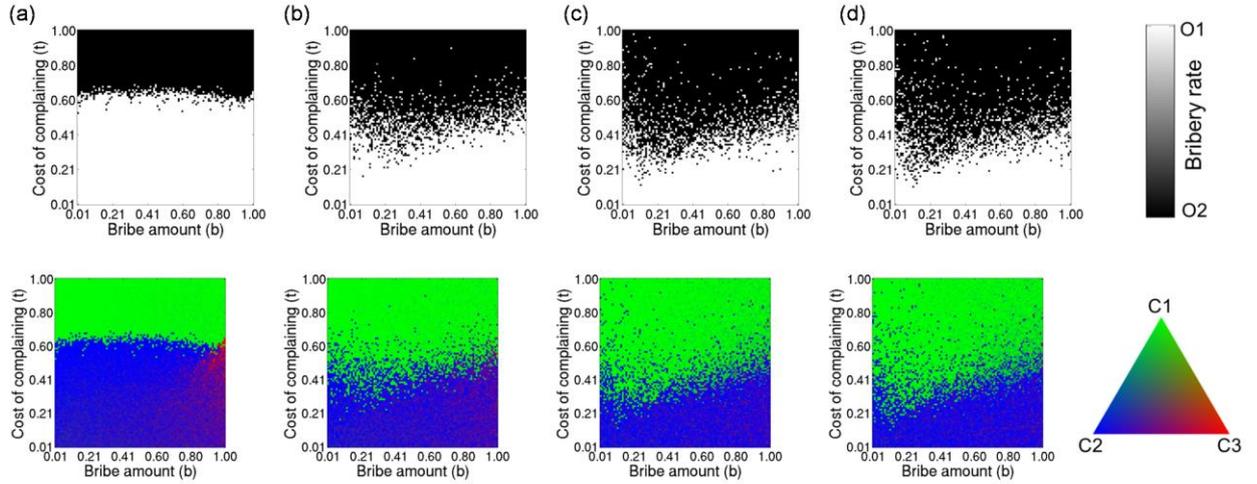

Supplementary Fig. 3: Equilibrium distribution for officers (upper panel) and citizens (lower panel) for increasing rewiring probability of both citizen's and officer's network with parameter $N_C = 100$, $N_O = 100$, $IN = 1$, $ON = 2$, (a) $p = p' = 0$ (b) $p = p' = 0.1$ (c) $p = p' = 0.4$ and (d) $p = p' = 1$. Other fixed parameters are $v = 1$, $c = 1$, $k = 0.4$, $p_o = 2$, $p_c = 0$, $r = b$, $\mu = 10^{-4}$. Increasing randomness of both citizen's and officer's network hinders the spread of honest officer in the population.

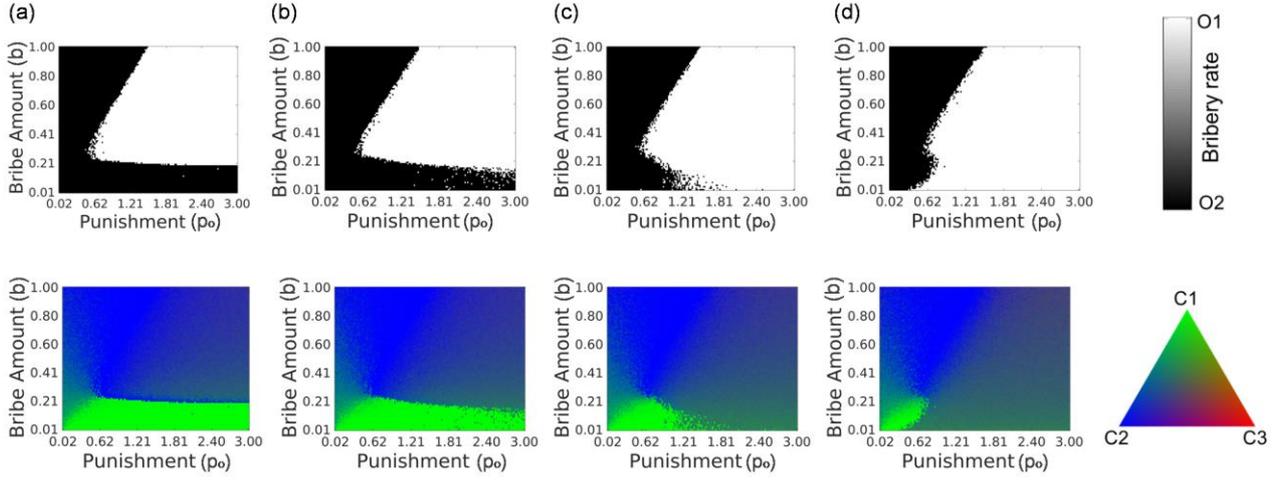

Supplementary Fig. 4: Equilibrium distribution of officers (upper panel) and citizens (lower panel) for different bribe-amount ($b$) and penalty imposed on corrupt officers ($p_o$). The rewiring probability of citizen's network corresponds to (a) $p = 0$ (regular) (b) $p = 0.1$ (c) $p = 0.4$ and (d) $p = 1.0$ (completely random). Fixed parameter in the simulation: $N_C = 1000$, $N_O = 100$, $IN = 1$, $CN = 6$, $ON = 2$, $v = 1$, $c = 1$, $k = 0.4$, $t = 0.1$, $p_c = 0$, $r = b$, $\mu = 10^{-4}$.

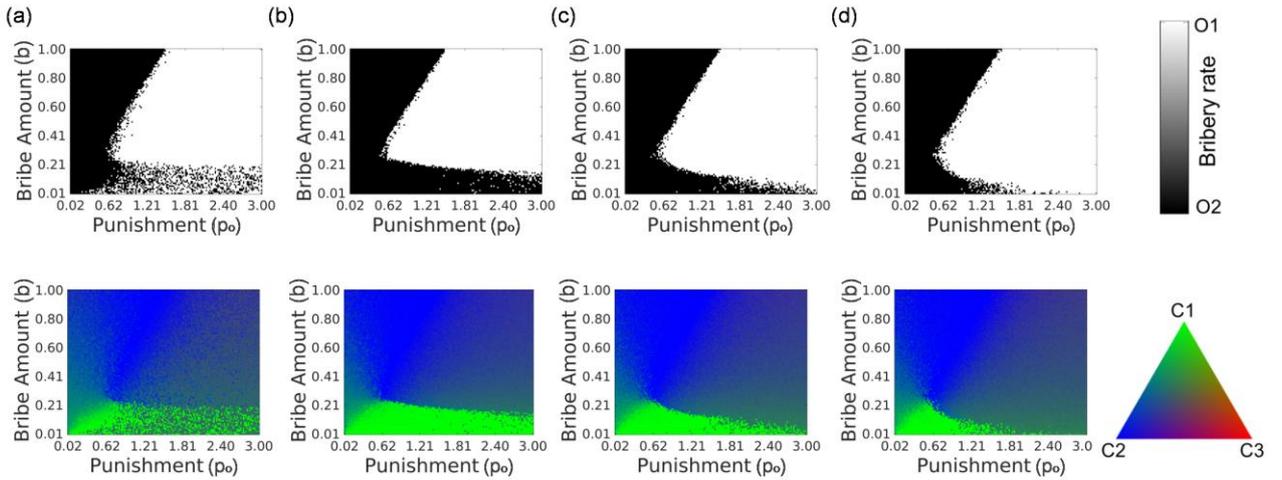

Supplementary Fig. 5: Equilibrium distribution for officers (upper panel) and citizens (lower panel) in the $b$-$p_o$ parameter space, showing the effect of increasing $CN$ with parameter $N_C = 1000$, $N_O = 100$, $IN = 1$, $ON = 2$, and $p = 0.1$ (a) $CN = 2$ (b) $CN = 6$ (c) $CN = 20$ and (d) $CN = 50$. Other fixed parameters in the simulation: $v = 1$, $c = 1$, $k = 0.4$, $t = 0.1$, $p_c = 0$, $r = b$, $\mu = 10^{-4}$.

Supplementary Movie 1: The videos show the evolution of strategies of citizens and officers in their respective networks over time. Populations of both citizens and officers are initialized with an equal number of randomly distributed players for each strategy. Black and white nodes represent corrupt and honest officers respectively. Green, blue and red nodes represent citizens who pay silently ($C_1$),

pay & complain ($C_2$) and refuse to pay ($C_3$) respectively. Officers are embedded on a network with regular linear chain topology with $ON = 2$. Citizens, on the other hand, are embedded in a network characterized by (a) $p = 0.1$, $CN = 2$ (b) $p = 1$, $CN = 2$ (c) $p = 0.1$, $CN = 20$ (d) $p = 1$, $CN = 20$. Each officer is also connected with a citizen directly below it with the degree of the interaction network being $IN = 1$. Other simulation parameters are fixed to the following values: $N_C = 50$, $N_O = 50$, $v = 1$, $c = 1$, $b = 0.2$, $t = 0.5$, $k = 0.4$, $p_o = 2$, $p_c = 0$, $r = b$.